\documentclass[aps, 
pra,
twocolumn,
superscriptaddress,
showpacs,preprintnumbers,amsmath,amssymb,
nofootinbib,
longbibliography,floatfix]{revtex4-2}
\usepackage[
colorlinks=true, 
pdfstartview=FitV, 
linkcolor=blue, 
citecolor=magenta, 
urlcolor=blue, 
bookmarks=true,
bookmarksnumbered=true,
pdftitle={},
pdfauthor={}
]{hyperref}

\usepackage[dvips]{graphicx}
\usepackage{bm,latexsym,amsmath,amssymb,amsfonts,mathrsfs,textcomp}
\usepackage{color}
\input{colordvi.tex}
\usepackage{comment}
\usepackage{simpler-wick}
\usepackage{tikz}
\usepackage[compat=1.1.0]{tikz-feynhand}
\usepackage[normalem]{ulem}
\usepackage{subfigure}

\newcommand{\T}{\mathrm{T}}

\newcommand{\diff}{\mathrm{d}} 
\newcommand{\rmi}{\mathrm{i}} 
\newcommand{\rme}{\mathrm{e}}

\newcommand{\hH}{\hat{H}}

\newcommand{\hpsi}{\hat{\psi}}

\newcommand{\ptc}[1]{{\bar{#1}}}

\newcommand{\hvarphi}{\hat{\varphi}}
\newcommand{\hPhi}{\hat{\Phi}}

\newcommand{\hn}{\hat{n}}

\newcommand{\pn}{\ptc{n}}

\newcommand\Lcal{\mathcal{L}}

\newcommand\Ocal{\mathcal{O}}

\newcommand{\average}[1]{\langle#1\rangle}

\renewcommand\Re{\mathop{\mathrm{Re}}}
\renewcommand\Im{\mathop{\mathrm{Im}}}

\newcommand{\bx}{\bm{x}}

\newcommand{\bz}{\bm{z}}
\newcommand{\br}{\bm{r}}
\newcommand{\bk}{\bm{k}}

\newcommand{\bq}{\bm{q}}

\newcommand{\bnab}{\bm{\nabla}}
\newcommand{\bzero}{\bm{0}}

\newcommand{\U}{\text{U}}

\newcommand{\rsout}[1]{{{\color[rgb]{1,0,0}{\sout{#1}}}}}
\newcommand{\cout}[1]{ \if 0 {#1} \fi }

\begin{document}
\title{Complex-valued in-medium potential between heavy impurities in ultracold atoms}

\author{Yukinao Akamatsu}
    \email{yukinao.a.phys@gmail.com}
    \affiliation{Department of Physics, Osaka University, Osaka 560-0043, Japan}
\author{Shimpei Endo}
    \email{shimpei.endo@uec.ac.jp}
  \affiliation{Department of Engineering Science, The University of Electro-Communications, Tokyo 182-8585, Japan}
    \affiliation{Department of Physics, Tohoku University, Sendai 980-8578, Japan}
\author{Keisuke Fujii}
    \email{k.fujii@phys.s.u-tokyo.ac.jp}
    \affiliation{Institut f\"ur Theoretische Physik, Universit\"at Heidelberg, D-69120 Heidelberg, Germany}
    \affiliation{Department of Physics, The University of Tokyo, 7-3-1 Hongo, Bunkyo-ku, Tokyo 113-0033, Japan}
\author{Masaru Hongo}
    \email{hongo@phys.sc.niigata-u.ac.jp}
    \affiliation{Department of Physics, Niigata University, Niigata 950-2181, Japan}
    \affiliation{RIKEN iTHEMS, RIKEN, Wako 351-0198, Japan}
    
\date{\today}

\begin{abstract}
We formulate the induced potential in a finite temperature cold atomic medium between two heavy impurities, or polarons, which is shown to be \textit{complex-valued} in general.  
The imaginary part of the complex-valued potential describes a decoherence effect, and thus, the resulting Schr\"odinger equation for the two polarons acquires a non-Hermitian term.
We apply the developed formulation to two representative cases of polarons interacting with medium particles through the $s$-wave contact interaction: 
(i) the normal phase of single-component (i.e., spin-polarized) fermions using the fermionic field theory, and 
(ii) a superfluid phase using the superfluid effective field theory, which is valid either for a Bose-Einstein condensate (BEC) of a single-component Bose gas or for the BEC-BCS crossover in two-component fermions at a low-energy regime.
Computing the leading-order term, the imaginary part of the potential in both cases is found to show a universal $r^{-2}$ behavior at long distance.
We propose three experimental ways to observe the effects of the universal imaginary potential in cold atoms.
\end{abstract}


\maketitle

\section{Introduction} \label{sec:1} 
The impurity problem is a paradigmatic quantum many-body problem, providing us with a fundamental aspect on the nature of strongly correlated quantum matters. 
Historically, the Anderson localization~\cite{PhysRev.109.1492} and the Kondo effect~\cite{kondo1964resistance} are well-celebrated examples of the quantum impurity problem. Another prototypical example of the impurity problem is the polaron; while it was originally proposed in the solid-state physics~\cite{landau1948effective}, dressed quasiparticle states universally appear around an impurity immersed in various quantum matters.
Unveiling the effective interactions between the impurities is pivotal to discovering and elucidating non-trivial many-body phenomena.
For instance, the Ruderman-Kittel-Kasuya-Yosida (RKKY) interaction~\cite{Ruderman-Kittel:1954,Katsuya:1956,Yoshida:1957}, acting between localized spins in solids, gives rise to a spin-glass phase.
In subatomic physics, the impurity problem has been used to distinguish different phases of matter.
Quark confinement and deconfinement are characterized by how gluon fields respond to heavy quarks, regarded as impurities~\cite{Wilson:1974sk, McLerran:1981pb}.
Despite the long history of impurity physics, several long-standing problems are often caused by the difficulty in finding a good basis to handle the impurity states and in-medium interaction between them, which is vital in the study of many-body physics of impurities.

Recently, new and promising avenues have been opened for exploring the impurity physics with unprecedented precision and degree of controls; in the field of cold atoms, a few impurity atoms immersed in low-temperature Fermi and Bose atomic gases have been successfully realized in cold atoms, referred to as Fermi polarons~\cite{PhysRevLett.102.230402,kohstall2012metastability,koschorreck:2012,PhysRevLett.118.083602,PhysRevLett.122.093401} and Bose polarons~\cite{PhysRevLett.117.055302,PhysRevLett.117.055301,PhysRevA.99.063607,Yan:2020,morgen2023quantum}, respectively.
The polaron problem has attracted renewed interests~\cite{PhysRevA.74.063628,Devreese_2009,Chevy_2010,Massignan2014}, especially after its one-particle properties, e.g., energies, lifetimes, residues, and effective masses, have accurately been observed for variable interaction strengths and temperatures in cold atom experiments~\cite{PhysRevLett.102.230402,kohstall2012metastability,koschorreck:2012,PhysRevLett.118.083602,PhysRevLett.122.093401,PhysRevLett.117.055302,PhysRevLett.117.055301,PhysRevA.99.063607,Yan:2020,morgen2023quantum}.

Beyond the one-particle properties, the exploration of medium-induced interactions between polarons has been an ongoing hot subject.
On the experimental side, the polaron-polaron interaction has been observed for a degenerate Fermi gas in a perturbative regime~\cite{desalvo2019observation}, and its signatures have also been observed in the non-perturbative regime~\cite{PhysRevLett.102.230402,PhysRevLett.118.083602,baroni2023mediated}, in conjunction with a precursor of the phase separation and the long-sought itinerant ferromagnetism~\cite{PhysRevLett.118.083602}. 
It is also notable that the polaron phenomena have recently been achieved in semiconductor materials, in which signatures of the polaron-polaron interaction have been observed for bosonic~\cite{tan2022bose} and fermionic systems~\cite{muir2022interactions}. 
On the theoretical side, the induced potential is found to show an oscillating behavior analogous to the RKKY interaction for the fermionic medium~\cite{PhysRevA.79.013629,Macneill2011,Endo2013,Enss2020}, while monotonic attractive interactions appear for the bosonic one~\cite{Pethick2008,PhysRevB.93.205144,naidon2018two,Guardian:2018a,Guardian:2018b,Panochko:2021,Panochko:2022,Fujii2022,Drescher:2023}. 
In particular, Ref.~\cite{Fujii2022} has found that a superfluid medium, irrespective of its microscopic details, should induce a universal long-range power-law potential between the polarons, showing novel universal feature of the polaron-polaron correlations.

Generally, however, such real-valued interaction potentials cannot describe all of the relevant medium effects.
Polaron physics is inherently an open quantum system problem, possessing non-Hermitian nature due to its environmental medium effect.
Recently, the open quantum system problem has been believed to be relevant in broad fields of science, ranging from quantum computations~\cite{preskill2018quantum,RevModPhys.94.015004}, topological quantum matters~\cite{bergholtz2021exceptional,PhysRevX.8.031079}, to photosynthetic biological materials~\cite{ishizaki2009unified,romero2014quantum}: coupling to an environment not only can lead to decoherence and dissipation of the system but also can bring about entirely new emergent phenomena.
To elucidate the open-system nature of polaron physics, it is essential to reconsider the medium-induced interaction, which is a fundamental building block of the many-body physics of polarons; due to the non-Hermitian character, it is naturally expected that the induced potential becomes {\it complex-valued}.
The complex-valued potential between the heavy quarks~\cite{Laine:2006ns,Beraudo:2007ky,Brambilla:2008cx,Rothkopf:2011db} have also been believed to be important in clarifying quark dynamics of quark-gluon plasmas---a hot QCD matter created during heavy-ion collision experiments.

As the real part of the polaron-polaron potential shows a universal feature~\cite{Fujii2022}, it is natural to ask the question: ``does the imaginary part of the polaron-polaron potential also show universality?" 
Answering this question not only contributes to elucidating the dynamics of the polarons in cold atoms and quarks in quark-gluon plasmas, but would unveil a novel universal aspect of the open-quantum system problem, especially in a highly non-trivial interacting quantum many-body setup.

To address this question, we study in this paper the complex-valued in-medium potential between the two heavy polarons in cold atomic systems, especially focusing on its imaginary part.
Following the definition developed in the context of hot QCD plasmas~\cite{Laine:2006ns,Beraudo:2007ky,Brambilla:2008cx,Rothkopf:2011db}, we first lay out a solid theoretical basis of the complex-valued in-medium potential suited for the study of cold atomic systems, which is applicable even in the case with strong impurity-medium coupling.
We also present a practical formula for the complex-valued potential when the impurity-medium coupling is weak, which was first derived in~\cite{sighinolfi2022stochastic} in the context of cold atomic gases using the influence functional formalism.
We then evaluate the imaginary part of the in-medium potential for polarons immersed in two representative examples of the finite-temperature quantum medium---a Fermi gas and a superfluid gas. Analogous to the real part, we identify that it reproduced an oscillating-decaying behavior for the Fermi polarons at low temperatures \cite{sighinolfi2022stochastic} while it exhibits a monotonic decay for the polarons in the superfluid.
Notably, we find the long-range asymptotic behavior shows the same power-law decay given by $V_{\mathrm{Im}} (r) \propto 1/r^2$ in both cases.
We stress that this is not a coincidence but rather a universal result: we show that the $1/r^2$ imaginary potential universally appears for any physical system if the collision between the polaron and an elementary excitation in the medium is almost elastic. 
We also discuss three experimental manifestations of the imaginary part of the complex-valued potential; that is, a direct detection through radio-frequency interferometry, and relatively indirect methods via spectral width of bipolaron, and relaxation of density fluctuation induced by the single impurity.

The paper is organized as follows. 
In Sec.~\ref{sec:definition}, we provide a general definition of the in-medium potential and formulae to evaluate the real and imaginary parts applicable when the impurity and medium is weakly coupled through a $s$-wave channel.
In Sec.~\ref{sec:calculating-ImV}, we evaluate the imaginary part of the in-medium potential acting on two polarons immersed in a non-interacting Fermi gas and a superfluid gas and discuss the physical origin of its universal $r^{-2}$ behavior.
In Sec.~\ref{sec:experiments}, we discuss experimental protocols to observe the imaginary potential.
Sec.~\ref{sec:conclusion} is devoted to the conclusion and future outlook.
In Appendices 
\ref{sec:weak-coupling}, \ref{sec:appendix2}, and \ref{sec:appendix1},
we provide detailed derivations of some results in the main text, and the relation between our work and the Langevin equation.
Throughout this paper, a natural unit $\hbar=k_B=1$ is used.

\section{In-medium potential} \label{sec:definition}
In this section, we lay out the basis of the in-medium potential between the two impurities immersed in a finite-temperature medium relying on the real-time Green's function, following Ref.~\cite{Laine:2006ns}.
After presenting a general definition of the in-medium potential, we give a formula for the weakly-coupled impurity-medium case reproducing the one derived in Ref.~\cite{sighinolfi2022stochastic}, which will be used in the subsequent sections.

Consider two impurities with their masses $M$ immersed in a finite-temperature Fermi or Bose medium.
To define the medium-induced potential acting on the impurities, we rely on an impurity two-body correlation function, which we can regard as a generalization of the wave function for the relative coordinates of the two impurities in the medium.
Using the impurity creation operator $\hat{\Phi}^{\dagger}(\bx)$ at position $\bx$, we introduce the following real-time correlation function:
\begin{align}
 \Psi(\br,t) = 
 \int \diff^3 R\,
 \big\langle
 &\hat{U}^{\dagger}(t,0) \hat{\Phi} (\bm{R} - \br/2 ) \hat{\Phi} (\bm{R}+\br/2) \hat{U} (t,0) 
 \nonumber \\ &\times 
 \hat{\Phi}^{\dag}(\bx_1) \hat{\Phi}^{\dag}(\bx_2)
 \big\rangle,
 \label{eq:impurity-Green-fcn}
\end{align}
where $\langle\cdots\rangle$ denotes the average over the density operator 
$\hat{\rho}_0 = \hat{\rho}_{\mathrm{eq}}^{\mathrm{med}}
\otimes |0_{\mathrm{imp}} \rangle \langle 0_{\mathrm{imp}}|$ 
with an equilibrium density operator for the medium $\hat{\rho}_{\mathrm{eq}}^{\mathrm{med}}$ and the zero-particle state for the impurity $|0_{\mathrm{imp}} \rangle$, and $\hat{U}(t,0)$ is a time-evolution operator of the total system from $0$ to $t$.
We let the initial condition, or $\bx_1$ and $\bx_2$ dependence, of the wave function implicit.
One can then confirm that $\Psi(\br,t)$ reduces to the usual wave function for the relative motion of two impurities separated by a distance $\bm{r}$ in the absence of the surrounding thermal medium.

Regarding $\Psi(\br,t)$ as a generalization of the wave function motivates us to introduce an in-medium potential.
In fact, one finds that $\Psi(\br,t)$ effectively obeys the Schr\"odinger equation at long time as 
\begin{align}
 \rmi \frac{\partial}{\partial t} \Psi(\br,t) 
 \simeq 
 \left[ - \frac{\bnab_r^2}{M_*} + \bar{V} (\br, \bnab_r) \right]
 \Psi(\br,t),
 \label{eq:2body-complex-energy}
\end{align}
where we introduced the energy of two impurities in the medium $\bar{V} (\br, \bnab_r)$ subtracted by the kinetic energy with an in-medium effective impurity mass $M_*$.
We note that $\bar{V} (\br, \bnab_r)$ generally contains spatial derivatives due to impurity motions.
In the limit where the impurity is sufficiently heavy, we can treat the impurity as a test particle almost fixed at a certain position, and thus drop the derivative dependence of the kinetic term and $\bar{V}(\br,\bnab_r)$.
As a result, the time-derivative and the potential only survive in Eq.~\eqref{eq:2body-complex-energy}.
We can then define the in-medium energy for two infinitely heavy impurities as follows:
\begin{equation}
 \bar{V} (\br)\equiv \lim_{t \to \infty} \lim_{M\to \infty}
 \frac{\rmi}{\Psi(\br,t) }
 \frac{\partial}{\partial t} \Psi(\br,t) .
 \label{eq:def-complex-Vbar}
\end{equation}
Equivalently, by matching the evaluated long-time behavior of $\Psi(\br,t)$ with $\Psi(\br,t) \simeq \rme^{- \rmi \bar{V} (\br) t}\Psi(\br,0)$, we can identify $\bar{V} (\br)$, which we will adopt in the next subsection. 

There is one small subtlety to regard Eq.~\eqref{eq:def-complex-Vbar} as a medium-induced interaction potential between the impurities; $\bar{V} (\br)$ does not vanish even when the two impurities are infinitely separated as $r\equiv |\br| \to \infty$.
This non-vanishing contribution results from the complex-valued one-body energy of a single heavy impurity immersed in the finite-temperature medium.
We thus define the in-medium genuine interaction potential $V(\br)$ by subtracting this remaining contribution as 
\begin{align}
 V (\br)\equiv  \bar{V} (\br)  - 2 E^{(1)}
 ~~\mathrm{with}~~
 2 E^{(1)} \equiv  \lim_{r \to \infty} \bar{V} (\br).
 \label{eq:def-complex-V}
\end{align}
As shown later, the presence of the finite-temperature medium makes the induced potential complex-valued as 
\begin{equation}
 V(\br) = V_{\mathrm{Re}} (\br) + \rmi V_{\mathrm{Im}} (\br) ,
\end{equation}
whose imaginary part causes the dissipation and decoherence in the impurities' dynamics.

According to Eqs.~\eqref{eq:def-complex-Vbar} and \eqref{eq:def-complex-V}, we need to evaluate the long-time behavior of $\Psi(\br,t)$ to find the in-medium potential $V(\br)$.
While the definition presented so far works in general, it is often difficult to accurately compute the impurity correlation function~\eqref{eq:impurity-Green-fcn}, particularly in the case of strong impurity-medium coupling.
Thus, assuming that the impurity-medium coupling is weak, we here provide a practically useful formula for the complex-valued potential in terms of the retarded Green's function of the medium.
This weak-coupling assumption is usually valid for polaron physics in cold atom experiments, as long as the impurity and medium atoms are away from any resonance and the impurity-medium $s$-wave scattering length is small.
We note that the interaction between the medium particles are not required to be weak, but can be strong, e.g. unitary gas (see Sec.~\ref{sec:superfluid}).

We now specify the total Hamiltonian of the system as 
\begin{equation}
 \hat{H}_{\mathrm{total}} = \hat{H}_{\mathrm{med}}  + \hat{H}_{\mathrm{int}} + \hat{H}_{\mathrm{imp}},
 \label{eq:H-tot}
\end{equation}
where $\hat{H}_{\mathrm{med}}$ and $\hat{H}_{\mathrm{imp}}$ give the Hamiltonian for the medium and impurity while $\hat{H}_{\mathrm{int}}$ describes the interaction between the impurity and the medium particles.
Throughout this paper, we assume that the medium and impurity particles interact through the $s$-wave contact (or density-density) interaction, as it can properly describe dominant low-energy scatterings in cold atoms. 
The interaction Hamiltonian is then given by
\begin{align}
 \hat{H}_{\mathrm{int}} = 
 g \int \diff^3 x\,
 \hat{\Phi}^\dagger(\bx)\hat{\Phi}(\bx)\hat{n}(\bx),
 \label{eq:H-int}
\end{align}
where $g$ is the impurity-medium coupling constant.
One can relate $g$ to the $s$-wave scattering length $a_{\mathrm{IM}}$ between the impurity and medium particles as $g = 2 \pi a_{\mathrm{IM}} \left( \frac{1}{m} + \frac{1}{M} \right)$ with masses $m$ and $M$ for the medium and impurity particles~\cite{braaten2006universality,pethick2008bose}.
We also introduced the particle number density operator for the medium $\hat{n}(\bx)$.

For a weak impurity-medium coupling $g$, the real-time correlation function $\Psi(\br,t)$ can be computed by the ladder approximation. Using $\Psi(\br,t) \simeq \rme^{- \rmi \bar{V} (\br) t}\Psi(\br,0)$ and Eq.~\eqref{eq:def-complex-V}, we obtain the complex-valued potential as follows (see Appendix \ref{sec:weak-coupling} for details):
\begin{align}
 V_{\mathrm{Re}} (\br) 
 &= -g^2 \lim_{\omega \to 0} G^R (\bm r, \omega),
 \label{eq:perturbative-V-Re}
 \\
 V_{\mathrm{Im}} (\br) 
  &= -  g^2 
  \lim_{\omega\to 0} \frac{2T}{\omega}
  \Im G^R(\bm r, \omega),
  \label{eq:perturbative-V-Im}
\end{align}
where we introduced the retarded Green's function 
of the number density
\begin{align}
 G^R(\bm r, \omega) 
 &\equiv \rmi \int_{-\infty}^{\infty} \diff t \,
 \rme^{\rmi\omega t} \theta (t) 
 \langle [
 \hat{n}(\bm r,t), 
 \hat{n}(\bm 0,0)]\rangle,
 \label{eq:retarded}
\end{align}
with $\hat{n}(\bm r,t)\equiv \rme^{\rmi \hat{H}_{\textrm{med}}t}\hat{n}(\bm r)\rme^{-\rmi \hat{H}_{\textrm{med}}t}$ and the Heaviside step function $\theta(t)$.
We also find the $\br$-independent one-body energy $E^{(1)} = E_{\Re}^{(1)} + \rmi E_{\Im}^{(1)}$ as
\begin{align}
 2 E_{\Re}^{(1)} 
 &= 2g \average{\hat{n}(\bzero)} + V_{\mathrm{Re}} (\br = 0) ,
 \\
 2 E_{\Im}^{(1)} 
 &= V_{\mathrm{Im}} (\br = 0) .
 \label{eq:imaginary-energy-disconnected}
\end{align}
We note that Eqs.~\eqref{eq:perturbative-V-Re} and \eqref{eq:perturbative-V-Im} reproduce the formula given in \cite{sighinolfi2022stochastic}, whose main subject was classical Langevin equation.
We provide a brief explanation in Appendix~\ref{sec:appendix2} on how the complex potential and the classical Langevin equation are related.

Some remarks are in order.
First, in the weak impurity-medium coupling regime, the in-medium potential $V(\br)$ is determined by the dynamical properties of the medium without the impurities [recall that the time-evolution of $\hat{n}(\bm x,t)$ is controlled by $\hH_{\mathrm{med}}$]. 
We will later use this to propose a way to experimentally investigate an imprint of the in-medium potential on medium dynamics with a single impurity (see Sec.~\ref{sec:medium-quench}).
Second, Eq.~\eqref{eq:perturbative-V-Im} is proportional to the temperature $T$.
As a result, the imaginary part representing the dissipative effect is peculiar to finite-temperature media.
Finally, as we show in the Appendix \ref{sec:weak-coupling}, the derived formula provides a natural generalization of the well-known formula at $T=0$.

\section{Evaluating in-medium potential}
\label{sec:calculating-ImV}

In this section, we consider two representative polaron systems realized in ultracold atom systems and evaluate the in-medium potential \eqref{eq:perturbative-V-Re} and \eqref{eq:perturbative-V-Im} by computing~\eqref{eq:retarded}:~impurities in a non-interacting Fermi gas in Sec.~\ref{sec:Fermi} and those in a superfluid in Sec.~\ref{sec:superfluid}. 
They correspond to Fermi and Bose polaron experiments realized in cold atoms~\cite{PhysRevLett.102.230402,kohstall2012metastability,koschorreck:2012,PhysRevLett.118.083602,PhysRevLett.122.093401,PhysRevLett.117.055302,PhysRevLett.117.055301,PhysRevA.99.063607,Yan:2020,morgen2023quantum}. 
These two examples exhibit a common power-law decay ($\propto r^{-2}$) in the imaginary part of the in-medium potential over long distances.
In Sec.~\ref{sec:long-range}, we discuss the origin behind this novel universal power-law decay, which is commonly shared by a wide class of physical systems.

\subsection{Polarons in Fermi gas}\label{sec:Fermi}
The polaron problem has attracted renewed interests~\cite{PhysRevA.74.063628,Devreese_2009,Chevy_2010,Massignan2014}
As a warm-up, let us start from a simple fermionic medium and show how the imaginary potential appears by microscopic processes.
An impurity immersed in a single-component non-interacting Fermi gas is called the Fermi polaron, which is realized as a single spin-up fermion in a Fermi sea of spin-down atoms in ultracold atoms~\cite{PhysRevLett.102.230402,kohstall2012metastability,koschorreck:2012,PhysRevLett.118.083602,PhysRevLett.122.093401}.
Let $\hpsi_f(\bx)$ be a field operator for a single-component fermion of the medium.
Then, the Hamiltonian of the system reads
\begin{equation}
 \hH
 = \int \diff^3 x 
 \left[
 \frac{1}{2m} |\bnab \hpsi_f|^2 
 - \mu |\hpsi_f|^2
 + g \hpsi_f^\dag \hpsi_f \hPhi^\dag \hPhi
 \right],
 \cout{
 \Lcal = 
 \psi^\dag 
 \left( 
  \rmi \partial_t + \frac{\bnab^2}{2m}
  + \mu
 \right) 
 \psi 
 - g \psi^\dag \psi \Phi^\dag \Phi.
 }
\end{equation}
where $m$ and $\mu$ denote the mass and chemical potential of the medium fermions.

\begin{figure}[t]
 \centering
 \includegraphics[width=0.45\linewidth]{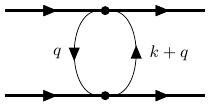}
 \caption{Feynman diagram representing the exchange of the particle-hole fluctuation (solid lines), which induces the potential at $O(g^2)$ between two impurities (amputated bold solid lines).
 }
 \label{fig:particle-hole-exchange}
\end{figure}

For non-interacting fermions, we can exactly compute the Green's function from the diagram in Fig.~\ref{fig:particle-hole-exchange} as
\begin{align}
 G (\bk, \rmi \omega_n^B)
 &=- \int_{\bq}
 \frac{n_F (\xi_{\bk+\bq}) - n_F (\xi_{\bq}) }{\rmi \omega_n^B -\xi_{\bk+\bq} + \xi_{\bq}},
\end{align}
with the bosonic Matsubara frequency $\omega^B_n = 2\pi n T$ at temperature $T$, Fermi distribution $n_F (E)\equiv 1/(\rme^{E/T} + 1)$, the energy measured from the Fermi sea $\xi_{\bk} = \frac{\bk^2}{2m} - \mu$, and a shorthand notations for the integral $\int_{\bk} \equiv \int \diff^3 k/(2\pi)^3$. Performing the analytic continuation and the Fourier transformation, the retarded Green's function is obtained as
\begin{align}
 G^R (\br,\omega) 
 =& - \int_{\bk,\bq}
 \rme^{\rmi (\bk - \bq ) \cdot \br}
 \frac{n_F (\xi_{\bk}) - n_F (\xi_{\bq}) }{\omega - \xi_{\bk} + \xi_{\bq} + \rmi 0}.  
\end{align}
Using Eqs.~\eqref{eq:perturbative-V-Re} and \eqref{eq:perturbative-V-Im}, we find 
\begin{align}
 V_{\Re} (\br)
 &= - g^2 \int_{\bk,\bq}
 \rme^{\rmi (\bk - \bq ) \cdot \br}
 \frac{n_F (\xi_{\bk}) - n_F (\xi_{\bq}) }{- \xi_{\bk} + \xi_{\bq}}. \\
 V_{\Im} (\br)
 &= - 2 g^2 T \int_{\bk,\bq}
 \rme^{\rmi (\bk - \bq ) \cdot \br}
 2 \pi \delta (\xi_{\bk} - \xi_{\bq}) 
 \nonumber \\
 &\hspace{40pt}
 \times
 \frac{1}{T} n_F (\xi_{\bk}) \big[ 1 - n_F (\xi_{\bk}) \big].  
\label{eq:complexpot-fermigas}
\end{align}
$V_{\Re} (\br)$ gives a finite-temperature real potential, which reduces to the RKKY interaction at $T=0$.
Our calculation reproduces the complex-valued potential in the same setup obtained by influence functional formalism \cite{sighinolfi2022stochastic}.

Performing the integrals for $V_{\Im} (\br)$, we obtain
\begin{equation}
 V_{\Im} (\br)
 = - \frac{8(k_F a_{\mathrm{IM}})^2 }{\pi}
 T f (2 k_F r, T/T_F),
 \label{eq:V-imaginary-Fermi}
\end{equation}
where we introduced 
the Fermi momentum $k_F\equiv \sqrt{2m\mu}$, Fermi temperature $T_F \equiv \mu$,
and the function $f (y,\tau)$ by 
\begin{equation}
 f (y,\tau) =
 \frac{1}{y} \int_0^\infty \diff s\,
 \frac{\sin s y}{\rme^{ (s^2 - 1)/\tau} + 1} .
 \label{eq:f-Fermi-polaron}
\end{equation}
In Eq.~\eqref{eq:V-imaginary-Fermi}, we also replaced the coupling constant $g \simeq 2\pi a_{\mathrm{IM}}/m$ relying on the heavy-impurity limit.

Expanding $f (y,\tau)$ with respect to $y$, we can evaluate asymptotic behaviors at $k_F r \ll 1$ or $k_F r \gg 1$ as 
\begin{align}
 \frac{V_{\Im} (r \ll k_F^{-1}) }{T_F}
 &\simeq - \frac{4(k_F a_{\mathrm{IM}})^2}{\pi} 
\left(\frac{T}{T_F}\right)^2
 \log (1 + \rme^{T_F/T} ),
 \label{eq:ImV-asymptotic-Fermi-short}
 \\
 \frac{V_{\Im} (r \gg k_F^{-1}) }{T_F}
 &\simeq - \frac{2(k_F a_{\mathrm{IM}})^2}{\pi} 
 \frac{T/T_F}{1+\rme^{-T_F/T}} \frac{1}{(k_F r)^2},
 \label{eq:ImV-asymptotic-Fermi-long}
\end{align}
where we neglected the higher-order correction with respect to $k_F r$ or $(k_F r)^{-1}$.
We note that the leading short-distance value in Eq.~\eqref{eq:ImV-asymptotic-Fermi-short} gives the imaginary part of the one-body energy [recall Eq.~\eqref{eq:imaginary-energy-disconnected}].
On the other hand, in the low-temperature limit, we obtain
\begin{equation}
 V_{\Im} (\br)
 \xrightarrow{T/T_F \ll 1} - \frac{8 (k_F a_{\mathrm{IM}} )^2}{\pi}
 T f(2 k_F r,0)
 \label{eq:ImV-low-T-limit-Fermi}
\end{equation}
with $f(y,0) = (1 - \cos y)/y^2$.
Similarly to the RKKY interaction, the imaginary part of the potential shows an oscillatory behavior in the low-temperature limit, whose period is controlled by the Fermi momentum $k_F$.
$V_{\Im} (\br)$ vanishes at $T=0$ because it is proportional to $T$.
Moreover, the exponent of the power-law decay is $V_{\Im} (\br) \propto r^{-2}$, which contrasts a difference with that for the RKKY interaction, $V_{\Re} (\br) \propto r^{-4}$.

\begin{figure*}
 \hspace{-4mm}
 \begin{minipage}{0.45\linewidth}
  \centering
  \includegraphics[width=0.95\linewidth]{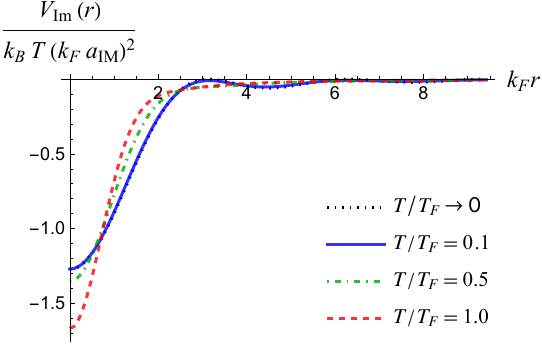} \\
 (a)
 \end{minipage}
 \begin{minipage}{0.45\linewidth}
  \centering
  \includegraphics[width=0.95\linewidth]{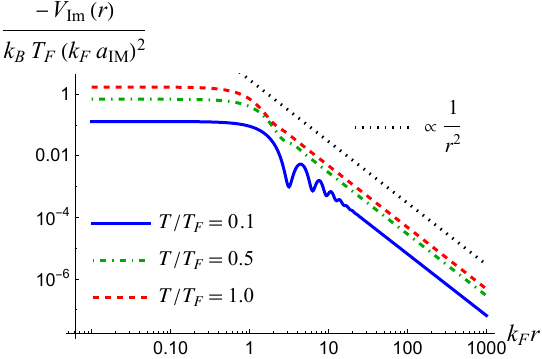} \\
  (b)
 \end{minipage}
 \caption{
 (a) The imaginary part of the induced potential for Fermi polarons with $T/T_F = 0.1,0.5$, and $1.0$ (solid blue, dashed-dotted green, and dashed red curves), normalized by $T$.
 The dotted black curve shows the zero-temperature limit of the imaginary potential.
 (b)  The imaginary part of the induced potential at long distances for Fermi polarons with  $T/T_F = 0.1,0.5$, and $1.0$ (solid blue, dashed-dotted green, and dashed red curves), normalized by $T_F$.
 The dotted black curve shows the power-law decay $r^{-2}$, which matches those of the induced potential.
 }
 \label{fig:Imaginary-potential-Fermi}
\end{figure*}

Figure \ref{fig:Imaginary-potential-Fermi} shows $V_{\Im} (\br)$ for several temperatures normalized by either $T_F$ or $T$.
The dashed curve in Fig.~\ref{fig:Imaginary-potential-Fermi} (a)---corresponding to the zero-temperature limit---demonstrates the oscillatory decay around $T=0$ against the inter-polaron separation $r$.
This oscillation is suppressed at higher temperature.
On the other hand, the magnitude of $V_{\Im} (\br)$ increases linearly as $T$.
This is because the imaginary part corresponds to the dissipative effect associated with the finite-temperature medium.
Figure~\ref{fig:Imaginary-potential-Fermi} (b) also demonstrates the $r^{-2}$ decay at long distances specified in Eq.~\eqref{eq:ImV-asymptotic-Fermi-long}.

\subsection{Polarons in superfluid gas}
\label{sec:superfluid}

As a second example, we consider impurities in superfluids.
Such a system is realized as impurities immersed in a BEC of the Bose gas~\cite{PhysRevLett.117.055302,PhysRevLett.117.055301,PhysRevA.99.063607,Yan:2020,morgen2023quantum}, or a pair-condensed Fermi superfluid including the unitary Fermi gas.

To clarify the long-range behavior of $V_{\Im}(r)$ in superfluid medium, we rely on an effective Hamiltonian of a superfluid phonon originally developed by Landau in describing the superfluid $^ 4$He~\cite{Khalatnikov,Lifshitz-Statphys2}. 
One advantage of this formulation is that it captures a universal long-distance behavior---guaranteed by the Nambu-Goldstone theorem~\cite{Nambu:1961tp,Goldstone:1961eq,Goldstone:1962es}---without relying on the perturbative treatment of the interaction between medium particles.
As a consequence, the following result covers not only the weakly-coupled Bose superfluid but also the strongly-coupled Fermi superfluid across BEC-BCS crossover, including the unitary regime, at the cost of the ability to describe the short-distance behavior around, e.g., a healing length. 
Our formulation is thus complementary to more familiar formulations for the Bose polarons such as the analysis based on the so-called Fr\"{o}hlich model~\cite{Frohlich:1954,Girardeau:1961}, or the variational wave functions~\cite{PhysRevA.74.063628,PhysRevLett.117.055302,PhysRevX.8.011024,PhysRevLett.117.113002}.

In the superfluid phase, we have the number density and phonon field as low-energy collective degrees of freedom as indicated by the spontaneously broken $\U(1)$ symmetry.
Let $\hn(\bx)$ and $\hvarphi(\bx)$ be such quantum operators, which satisfy the canonical commutation relation $[\hat{n}(\bx),\,\hat{\varphi}(\bx^\prime)]=- \rmi \delta^{(3)}(\bx-\bx^\prime)$.
We then find the effective Hamiltonian of the superfluid phonon interacting with the impurity as follows~\cite{Khalatnikov,Lifshitz-Statphys2}:
\begin{equation}
 \hH_{\mathrm{eff}}
 = \int \diff^3 x\,
 \left[
 \frac{\hn}{2m}  (\bnab \hvarphi)^2
 + \epsilon (\hn)
 + g \hn \hPhi^\dag \hPhi
 \right],
 \label{eq:Heff-superfluid}
\end{equation}
where $m$ is the mass of medium particles and $\epsilon (\hn)$ denotes an internal energy as a function of the number density.
\footnote{
The effective Hamiltonian~\eqref{eq:Heff-superfluid} is equivalent to the effective Lagrangian employed in~\cite{Fujii2022}, from which the real part of the in-medium potential was derived.
On the equivalence between these, we refer to Ref.~\cite{Son:2005rv}.
}

Considering the fluctuation around the global equilibrium, we expand $\hn (\bx) = \bar{n} + \delta \hn (\bx)$ and expand $\epsilon (\hn)$ at the quadratic order.
The effective Hamiltonian then reads
\begin{align}
 \hH_{\mathrm{eff}}
 \simeq \int \diff^3 x\,
 \biggl[
 &
 \frac{\pn}{2m} (\bnab \hvarphi)^2
 + \frac{1}{2\chi} (\delta \hn )^2
 + \frac{1}{2m} \delta \hn (\bnab \hvarphi)^2
 \nonumber \\
 &+ g \pn \hPhi^\dag \hPhi
 + g \delta\hn \hPhi^\dag \hPhi
 \biggr],
 \label{eq:Heff-superfluid-expansion}
\end{align}
where we defined the inverse charge susceptibility $\chi^{-1} = \epsilon'' (\pn)$ and omit the constant term $\epsilon (\pn)$ and the higher-order term with more than three $\delta \hn$.
Since we are interested in the long-distance behavior, we can regard the interaction term appearing in the first line of Eq.~\eqref{eq:Heff-superfluid-expansion} as a perturbation. 
We emphasize that this treatment follows from the derivative expansion~\cite{Son:2005rv}, and does not require a small coupling constant among medium particles.

To evaluate the retarded Green’s function for the number density operator $\hat{n}(\bx)$, we employ again the imaginary-time formalism:
\begin{align}
 \Delta_{\varphi\varphi} (\bk,\rmi \omega^B_n)
 &= \parbox{2.0cm}{\vspace{0pt}\includegraphics[width=2.0cm]{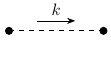}}
 = \frac{\chi^{-1}}{(\omega_n^B)^2 + E_{\bk}^2},
 \nonumber \\
 \Delta_{nn} (\bk,\rmi \omega^B_n)
 &= \parbox{2.0cm}{\vspace{0pt}\includegraphics[width=2.0cm]{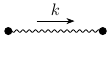}}
 = \frac{\chi E_{\bk}^2}{(\omega_n^B)^2 + E_{\bk}^2},
 \nonumber \\
 \Delta_{\varphi n} (\bk,\rmi \omega^B_n)
 &= \parbox{2.0cm}{\vspace{0pt}\includegraphics[width=2.0cm]{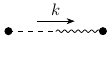}}
 = \frac{\omega_n^B}{(\omega_n^B)^2 + E_{\bk}^2},
 \nonumber \\
 \Delta_{n\varphi} (\bk,\rmi \omega^B_n)
 &= \parbox{2.0cm}{\vspace{0pt}\includegraphics[width=2.0cm]{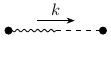}}
 = \frac{-\omega_n^B}{(\omega_n^B)^2 + E_{\bk}^2},
\end{align}
where we defined $E_{\bk}\equiv c_s |\bk|$.
From Eq.~\eqref{eq:Heff-superfluid-expansion}, we also read off the following three-point interaction vertex joining two phonon and one number density fluctuations:
\begin{equation}
\parbox{1.8cm}{\vspace{0pt}\includegraphics[width=1.8cm]{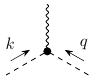}}
 = \frac{1}{2m} \bk \cdot \bq.
 \label{eq:medium-interaction-superfluid}
\end{equation}

\begin{figure}[t]
 \centering
 \includegraphics[width=0.95\linewidth]{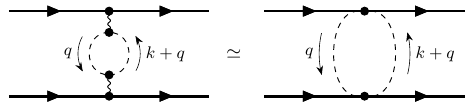}
 \caption{Feynman diagram representing the exchange of the phonon (solid lines), which induces the potential between two impurities (amputated bold solid lines) at $O(g^2)$.
 At low frequencies, the left diagram can be effectively reduced to the right one (see the main text).
 }
 \label{fig:phonon-exchange}
\end{figure}

We now perform a perturbative expansion with respect to the medium interaction term~\eqref{eq:medium-interaction-superfluid}: the leading one-loop diagram that contributes to the in-medium potential is given in the left diagram of Fig.~\ref{fig:phonon-exchange}. Recalling Eqs.~\eqref{eq:perturbative-V-Re} and \eqref{eq:perturbative-V-Im}, we evaluate the low-frequency behavior of the retarded Green's function.
At vanishing frequencies, the propagation of the density fluctuation $\delta\hat{n}(\bx)$ is suppressed as $\Delta_{nn} (\bk,\rmi \omega^B_n=0) = \chi$.
Thereby, the Green's function of the number density directly connected to the impurity lines on the left diagram of Fig.~\ref{fig:phonon-exchange} can be replaced with $\Delta_{nn} (\bk, \omega) \simeq \chi$ after analytic continuation to a low real frequency.
The resulting diagram is shown on the right of Fig.~\ref{fig:phonon-exchange}, which agrees with the diagram computed in Ref.~\cite{Fujii2022} to evaluate the real part of the potential. 

We thus find the corresponding Green's function as
\begin{widetext}
\begin{align}
 G (\bk, \rmi \omega_n^B)
 &\simeq 
 \Delta_{nn} (\bk, \rmi \omega_n^B)
 + \frac{\chi^2}{2m^2} 
  \int_{\bq} [\bq \cdot (\bk + \bq)]^2
  T \sum_l
 \Delta_{\varphi\varphi} (\rmi \omega_n^B + \rmi \omega_l^B, \bk + \bq) 
 \Delta_{\varphi\varphi} (\rmi \omega_l^B,\bq)
 \nonumber \\
 &\simeq \chi
 - \frac{1}{2m^2} 
 \int_{\bq} \frac{[\bq \cdot (\bk + \bq)]^2}{4 E_{\bk+\bq} E_{\bq}}
 \Bigg[ 
  \big[ 1 + n_B (E_{\bk+\bq}) + n_B (E_{\bq}) \big]
  \left( 
   \frac{1}{\rmi \omega_n^B - E_{\bk+\bq} - E_{\bq}}
   - \frac{1}{\rmi \omega_n^B + E_{\bk+\bq} + E_{\bq}}
  \right)
 \nonumber \\
 &\hspace{130pt}
 + \big[ n_B (E_{\bk+\bq}) - n_B (E_{\bq}) \big]
  \left( 
   \frac{1}{\rmi \omega_n^B + E_{\bk+\bq} - E_{\bq}}
   - \frac{1}{\rmi \omega_n^B - E_{\bk+\bq} + E_{\bq}}
  \right)
 \Bigg],
\end{align}
\end{widetext}
where we also added the tree-level contribution as the first term.
Here, $n_B (E)\equiv 1/(\rme^{\beta E} - 1)$ is the Bose distribution.
Performing the analytic continuation and the Fourier transformation, we find the retarded Green's function $G^R (\br, \omega)$, from which the in-medium potential is obtained.
The real part agrees with that computed in Ref.~\cite{Fujii2022} (except for the contact interaction arising from the first term), while we identify the imaginary part as
\begin{align}
 V_{\Im}(\bm r) &= 
 - \frac{2\pi g^2}{m^2} \int_{\bk,\bq} 
 \rme^{\rmi(\bk-\bq) \cdot \br} 
 \frac{(\bq\cdot\bk)^2}{4 E_{\bk}^2}
 \delta ( E_{\bk}- E_{\bq}) \nonumber \\ 
&\quad \times 
 [1+n_B(E_{\bk})] n_B (E_{\bk}) .
 \label{eq:impot-superfluid}
\end{align}
Performing the angular integration with $g \simeq 2\pi a_{\mathrm{IM}}/m$, we obtain the following result (see Appendix.~\ref{sec:appendix1}):
\begin{align}
 V_{\Im} (r) 
 &=  - \frac{a_{\mathrm{IM}}^2 T^7}{2\pi m^4 c_s^{10}} h (k_T r)
 \label{eq:impot1}
\end{align}
where $k_T \equiv T/c_s$ is a thermal phonon momentum and
\begin{align}
 h (y) &\equiv 
 \int_0^\infty \diff s\,
 \frac{ s^6\rme^{s}}{(\rme^s-1)^2} \nonumber \\
 &\quad \times 
 \left[
  \frac{3j_1 (sy)^2 }{(sy)^2} 
  - \frac{2j_1 (sy) j_2 (sy)}{sy} 
  + j_2 (sy)^2
 \right],
 \label{eq:h-polaron-superfluid}
\end{align}
with the spherical Bessel functions $j_n(x)$.

Using the asymptotic behavior of $h(y)$ (see Appendix.~\ref{sec:appendix1}), we find that $V_{\Im} (r) $ behaves at short and long distances (compared to a thermal phonon length scale $k_T^{-1}=c_s/T$) as
\begin{align}
 \frac{V_{\Im} (r \ll k_T^{-1} )}{mc_s^2}
 &\simeq  
 - \frac{8 \pi^5 a_{\mathrm{IM}}^2 T^7}{63 m^5 c_s^{12}}
 \label{eq:impot_asymp_short} 
 \\
 \frac{V_{\Im} (r \gg k_T^{-1} )}{mc_s^2}
 &\simeq  - \frac{\pi^3 a_{\mathrm{IM}}^2 T^5}{15 m^5 c_s^{10}} \frac{1}{r^2}
 \label{eq:impot_asymp_long}
\end{align}
According to Eq.~\eqref{eq:imaginary-energy-disconnected}, the leading short-distance value in Eq.~\eqref{eq:impot_asymp_short} gives the imaginary part of the energy of a single polaron.
Note that the imaginary part $V_{\Im} (r \gg k_T^{-1]})$ again shows a power-law decay at long distances $V_{\Im} (r \gg k_T^{-1}) \propto 1/r^2$, which is slower than the decay of the real part $V_{\Re} (r \gg k_T^{-1}) \propto 1/r^6$ specified in Ref.~\cite{Fujii2022}.

\begin{figure*}
 \hspace{-4mm}
 \begin{minipage}{0.45\linewidth}
  \centering
  \includegraphics[width=0.95\linewidth]{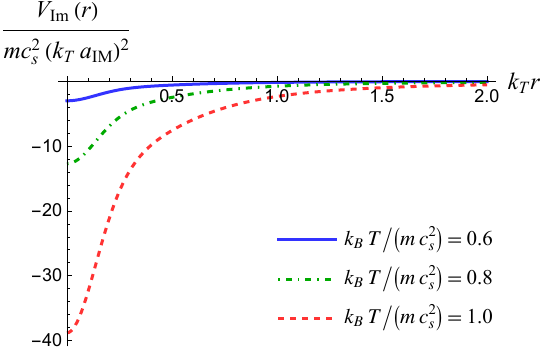} \\
 (a)
 \end{minipage}
 \begin{minipage}{0.45\linewidth}
  \centering
  \includegraphics[width=0.95\linewidth]{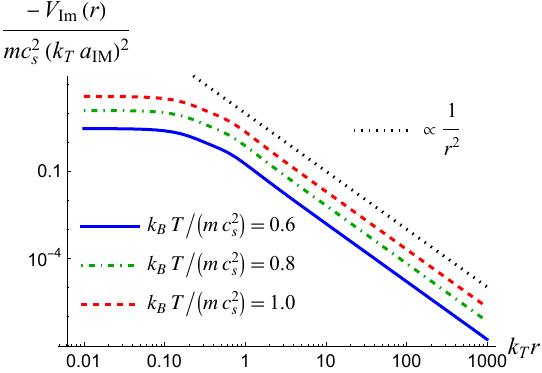} \\
  (b)
 \end{minipage}
 \caption{
 The imaginary part of the induced potential for polarons in the superfluid.
 (a) at short distances and 
 (b) long distances.
 Solid blue, dashed-dotted green, and dashed red curves show results for $T/(mc_s^2) = 0.6,0.8$, and $1.0$.
 In panel (b), the dotted black curve shows the power-law decay $r^{-2}$, which matches the asymptotic behaviors of the induced potential.
 }
 \label{fig:Imaginary-potential-Bose}
\end{figure*}

Figure~\ref{fig:Imaginary-potential-Bose} shows $V_{\Im}(r)$ in superfluids for several temperatures.
In contrast to the Fermi polaron case, there is no oscillatory behavior and the magnitude of the potential is more sensitive to the temperature.
One also explicitly confirms the $r^{-2}$ scaling at long distances, as clearly shown in the right panel of Fig.~\ref{fig:Imaginary-potential-Bose}.

\subsection{Universal long-range power-law decay}
\label{sec:long-range}

In the previous subsections, we have found that the imaginary part of the in-medium potential between polarons in non-interacting Fermi gases and superfluids both shows a power-law decay $V_{\Im} (r) \propto r^{-2}$ at long distances.
The real part of the potential similarly exhibits long-range power-law decay, which is attributed to the gapless nature of the mediated excitations, such as particle-hole excitations in Fermi gases and phonons in superfluids.
Accordingly, it is tempting to expect that the long-range power-law behavior of $V_{\Im}(r)$ is also due to the presence of gapless excitations.

However, this statement is \textit{not} true.
Indeed, there is an example in subatomic physics showing the same power-law decay $V_{\Im}(r) \propto r^{-2}$ at long distances without gapless excitations: the in-medium heavy-quark potential in hot QCD plasmas shows the same power law~\cite{Laine:2006ns,Beraudo:2007ky,Brambilla:2008cx} although the (electric) gluon acquires Debye screening mass bringing about the exponential decay for the real potential.
Namely, the gapless nature of the medium excitations is unnecessary.

The power-law behavior originates from the common structure of the low-energy scattering between the impurity and the medium excitation.
To clarify this origin, let us consider the imaginary potential in momentum space
\begin{align}
 \widetilde{V}^F_{\Im} (\bk) &\propto  
 - g^2 \int_{\bq} \hspace{-2pt}
 \delta (\xi_{\bq+\bk} - \xi_{\bq}) 
n_F (\xi_{\bq}) \big[ 1 - n_F (\xi_{\bq}) \big], \\
 \widetilde{V}^{\mathrm{SF}}_{\Im}(\bk) &\propto 
 - \frac{g^2}{m^2} \int_{\bq} \frac{(\bq^2 + \bk\cdot\bq)^2}{E_{\bq}^2}
 \delta ( E_{\bq+\bk}- E_{\bq}) \nonumber \\
 &\hspace{60pt} \times  n_B (E_{\bq})[1+n_B(E_{\bq})],
\end{align}
which follow from Eqs.~\eqref{eq:complexpot-fermigas} and \eqref{eq:impot-superfluid}, respectively.
The asymptotic behavior at long distances is then specified by that at low momentum regions $\bk \simeq \bzero$.

\begin{figure}[t]
 \centering
 \includegraphics[width=0.98\linewidth]{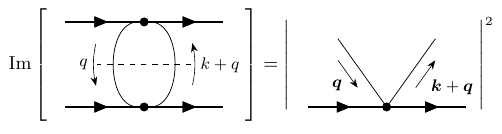}
 \caption{
 Scattering interpretation of the imaginary part of the induced potential.
 Taking the imaginary part results in cutting the diagram with the dashed line in the left panel.
 The imaginary part is thus given by the right panel, which gives a cross section of the on-shell medium excitation (solid lines) scattered by the static impurity (amputated bold lines).
 }
 \label{fig:scattering}
\end{figure}

These equations allow a scattering interpretation:
The imaginary part is expressed in terms of the scattering amplitude of the medium excitation with the static impurity as shown in Fig.~\ref{fig:scattering}~~(see, e.g., Ref.~\cite{LeBellac2000} for the so-called cutting rule). 
The integrand is composed of three parts; (a) the delta function imposing the energy conservation of the on-shell medium excitation, (b) the (properly normalized) scattering cross section part,\footnote{
Note that $E_{\bq}^{-2}$ appears for the superfluid case from the proper normalization with $\delta ( E_{\bq+\bk}- E_{\bq})$.}
and (c) thermal distributions for incoming and outgoing excitations.

Let us now investigate the low-$\bk$ behaviors of these building blocks.
Because the thermal distributions are $\bk$-independent, the low-momentum behavior is controlled by the delta function and cross section.
At low momentum $\bk \simeq \bzero$, one can approximate the delta functions as
\begin{align}
 \delta (\xi_{\bq+\bk} - \xi_{\bq}) &\to \delta (\bq\cdot\bk/m) = \frac{m}{qk}\delta(\cos\theta_{\bq}),
 \label{eq:delta-Fermi}
 \\
 \delta ( E_{\bq+\bk}- E_{\bq}) &\to \delta(c_s\bq\cdot\bk/q) = \frac{1}{c_s k}\delta(\cos\theta_{\bq}),
 \label{eq:delta-superfluid}
\end{align}
yielding the $k^{-1}$ factor ($\theta_{\bq}$ is the angle between $\bq$ and $\bk$, which is fixed $\theta_{\bq}=\pi/2$ by performing the $\bq$-integration).
As a result, if the low-energy scattering cross section approaches to the non-vanishing constant, the above $k^{-1}$ factor governs the long-distance behavior, leading to $r^{-2}$ behavior of $V_{\Im}(r)$ in coordinate space.

As the above discussion clarifies, the crucial point here is whether the low-energy scattering cross section gives the finite nonvanishing contributions at the vanishing momentum $\bk\to \bzero$.
The presence of nonzero contributions means that on-shell medium excitation can scatter off an immobile impurity with zero energy transfer, i.e., $(E_{\bq}, \bm q) \to (E_{\bq}, \bq + \bk)$, at low momentum $\bk \simeq \bzero$ (see the right panel of Fig.~\ref{fig:scattering}).
Indeed, the interaction vertices between the impurity and the medium excitation investigated in previous subsections have finite contributions in the limit of $\bk\to\bzero$, allowing for such scattering.

Therefore, when the impurity and the medium excitations can collide elastically, the imaginary potential shows a universal power-law decay, $V_{\Im}(r) \propto r^{-2}$, at long distances. 
Such a condition is expected to hold true not only for the Fermi and Bose polaron systems presented above, but rather for a wide variety of physical systems. 
In fact, the example in subatomic physics---the in-medium heavy-quark potential mentioned above---fits into this class.
On the other hand, the elastic scattering condition may be violated for not so heavy impurity, where the recoil effects are relevant~\cite{RoschQimp}.

\section{Experimental manifestation}
\label{sec:experiments}

In this section, we discuss how the imaginary potential between the heavy polarons can be observed in the experiments. Recently, the polaron has been realized in degenerate Fermi~\cite{PhysRevLett.102.230402,kohstall2012metastability,PhysRevLett.118.083602,PhysRevA.74.063628} and Bose~\cite{PhysRevLett.117.055302,PhysRevLett.117.055301} environments in cold atoms.
With high degree of controllability, the cold atom systems are expected to be the prime candidate to clearly observe the polaron-polaron interaction. Indeed, some signatures of the polaron-polaron interaction have already been observed in the experiments~\cite{desalvo2019observation,PhysRevLett.102.230402,PhysRevLett.118.083602,baroni2023mediated}. 
We discuss below how to experimentally observe the universal imaginary potentials between the heavy polarons in cold atoms.

\subsection{Radio-frequency interferometry signal}
\label{sec:echo}
One way to directly probe the spatial dependence of the polaron-polaron interaction $V(\bm{r})$ is based on the spectroscopic method used in Refs.~\cite{Cetina2016,Skou2021} to observe and characterize single polaron properties.
In Refs.~\cite{Cetina2016,Skou2021}, two sequences of radio-frequency pulses are applied to cold atoms, which generate a quantum superposition between the non-interacting and interacting states of the impurity particle with its surrounding atoms.
From the interference signal, we can read off a real-time single-particle Green's function of the impurity.

A similar approach can be used to observe $V(\bm{r})$ from the real-time Green's function~\eqref{eq:impurity-Green-fcn}: suppose that we have two impurities at fixed positions $\bm{R}_1$ and $\bm{R}_2$ in the non-interacting internal state with the surrounding atoms.
If we apply two sequences of radio-frequency pulses in the same manner as above, each impurity realizes a superposition of interacting and non-interacting states with its surroundings, and interferences occur between these states.
The interference between the states in which both the impurity interacts with its surroundings depends explicitly on $|\bm{R}_1-\bm{R}_2|$, while the others do not.
Hence, from the $|\bm{R}_1-\bm{R}_2|$-dependent signal, we can obtain the real-time Green's functions of the two impurities, from which the in-medium potential $V(\bm{R_1}-\bm{R_2})$ can be directly probed. 
The control of the impurity distance can be achieved in cold atoms by pinning the positions of the impurity atoms either via focused optical tweezers~\cite{schlosser2001sub,kaufman2021quantum}, or via a species-selected optical lattice~\cite{PhysRevA.75.053612,PhysRevLett.124.203201} which strongly traps the impurity atoms while leaving the surrounding environment atoms intact. 
Alternatively, if the observation sequence is much faster than the time scale of the heavy polaron's dynamics~\cite{Cetina2016,desalvo2019observation}, we can essentially neglect the motion of the impurity particles during the above observation sequence: in this case, $|\bm{R}_1-\bm{R}_2|$ can be controlled by varying the density of impurity atoms.

\subsection{Spectral width of bipolaron at low temperature}
\label{sec:lifetime}

The imaginary potential between the polarons affects the spectral width of the bound state of two polarons, i.e., bipolaron.
The imaginary part of the in-medium potential is smaller than the real part at low temperatures because it is proportional to the temperature as in Eq.~\eqref{eq:perturbative-V-Im}.
We can thus essentially regard the relative wave function of the bipolaron $\Psi_b(\br)$ created by the attractive $V_{\Re}(\br)$, and perturbatively estimate its spectral width $-\frac{1}{2}\Gamma$ as (see Appendix~\ref{sec:appendix2} for a more formal discussion)
\begin{align}
 \frac{1}{2}\Gamma 
 &\simeq - \int \diff^3 r \left|\Psi_b(\br)\right|^2 \bar{V}_{\Im} (\bm{r})
 \nonumber \\
 &= - 2E^{(1)}_{\Im} 
 - \int \diff^3 r \left|\Psi_b(\br)\right|^2 V_{\Im} (\bm{r}).
 \label{eq:width}
\end{align}
The width has contributions from both $V_{\Im}(\bm{r})$ and the imaginary part of the single-particle energy $2E^{(1)}_{\Im}$ (i.e. disconnected diagrams), which are collectively expressed using $\bar{V}_{\Im}(\bm{r})$, as in the first line.

Let us crudely evaluate the spectral width.
Because $V_{\Im}(\br)$ has its maximum absolute value at $\br = \bm 0$, the second term in Eq.~\eqref{eq:width} achieves that maximum $V_{\Im}(\bzero)$ when the two impurities get close.
Using Eq.~\eqref{eq:imaginary-energy-disconnected}, the width at that time is found as $\frac{1}{2}\Gamma \simeq - 4E^{(1)}_{\Im}$.
This result can be understood physically as follows:
The medium recognizes the two polarons as a single molecule with its coupling constant $2g$.
The imaginary part of the self-energy of this single molecule is given by that of one polaron with $g^2$ replaced by $(2g)^2$, resulting in the width of one molecule being four times that of one polaron.
In the opposite case where the two polarons are far apart, the second term in Eq.~\eqref{eq:width} is negligible compared to the first term, and thus the width is given by the sum of the contributions of the two independent polarons as $\frac{1}{2}\Gamma\simeq -2E^{(1)}_{\Im}$.
These two extreme cases give crude upper and lower bounds for the width as $2|E^{(1)}_{\Im}| < \frac{1}{2}\Gamma < 4|E^{(1)}_{\Im}|$, where the explicit form of $2E^{(1)}_{\Im} = V_{\Im}(\bm0)$ is found in Eqs.~\eqref{eq:ImV-asymptotic-Fermi-short} and \eqref{eq:impot_asymp_short}.
This inequality holds true as long as the imaginary part can be treated perturbatively.

The width of the two polarons can be observed in cold atoms by the radio-frequency spectroscopy~\cite{PhysRevLett.102.230402,kohstall2012metastability,PhysRevLett.118.083602,PhysRevLett.117.055302,PhysRevLett.117.055301}, with the rf pulse transferring the two impurities from a non-interacting initial state to an interacting final state.
It can also be measured from the decay rate of its Rabi oscillation~\cite{kohstall2012metastability,PhysRevLett.118.083602}, or via the width of a similar transition involving two polarons. 
It can also be observed from the dynamical behavior of the two impurities although it may be more challenging than the formers.

\subsection{Medium density fluctuation induced by a single impurity}
\label{sec:medium-quench}

Even with just a single impurity, we can also observe a signature of the universal imaginary potential between two polarons.
The idea is similar to what we learn in electromagnetism: we can observe an electric field induced by a single test charge, rather than a force acting from one test charge to another test charge.
In the present setup, as in Eq.~\eqref{eq:H-int}, the medium number density mediates the polaron-polaron interaction.
Thus, we can observe the complex-valued potential as the change in the medium number density resulting from the placement of a single test impurity in the medium.

The fluctuation of the medium number density is evaluated with the linear response theory, valid for our weak impurity-medium coupling setup, as
\begin{align}
 \average{\delta n (\bx,t)} 
 = - \int_{-\infty}^\infty \diff t' \int \diff^3 x'\,
 &G^R (\bx-\bx',t-t') 
 \nonumber \\
 &\times  A_0^{\mathrm{eff}} (\bx',t').
 \label{eq:linear-response-delta-n}
\end{align}
where the conjugate field $A_0^{\mathrm{eff}} (\bx,t)$ coupled to the medium number density is $A_0^{\mathrm{eff}} (\bx,t) = g(t)\hat{\Phi}^\dag (\bx,t) \hat{\Phi}(\bx,t)$ as read off from Eq.~\eqref{eq:H-int}.
While the above formula can be applied to various spatio-temporal dependence which may experimentally be explored in cold atoms, we here focus on the simplest case of a quench: we consider suddenly placing a heavy test impurity at the origin $\bx = \bzero$ at time $t=0$ as $A^{\mathrm{eff}}_{0}(\bx,t)=g\theta(t)\delta^{(3)}(\bx)$.
This is experimentally achieved in cold atom experiments by transferring the impurity atom from a non-interacting state to an interacting state with a rf pulse~\cite{kohstall2012metastability,PhysRevLett.118.083602,PhysRevLett.117.055302,PhysRevLett.117.055301}.
The long-time behavior of the medium number density after the quench is governed by the low-frequency behavior of $G^R (\bx-\bx',\omega)$.
Recalling Eqs.~\eqref{eq:perturbative-V-Re} and \eqref{eq:perturbative-V-Im}, we find
\begin{align}
  G^R (\bx, \omega)
 &= -\frac{1}{g^2}
 \left[
 V_{\Re} (\bx) + \rmi \frac{\omega}{2T}V_{\Im} (\bx) + O (\omega^2)
 \right]
 \nonumber \\
 &\approx
 -\frac{1}{g^2}
 \frac{2\rmi T \frac{V_{\Re} (\bx)^2}{V_{\Im} (\bx)}}{\omega + 2 \rmi T \frac{V_{\Re} (\bx)}{V_{\Im} (\bx)} } .
\end{align}
Extrapolating this low-frequency behavior and substituting it into Eq.~\eqref{eq:linear-response-delta-n}, we obtain 
\begin{equation}
 \begin{split}
 \average{\delta n (\bx,t)}
 &\approx 
 \frac{1}{g} \left[1 - \rme^{- 2T \frac{V_{\Re} (\bx)}{V_{\Im} (\bx)} t} \right]
 V_{\Re} (\bx).
 \end{split}
\end{equation}
Thus, the medium number density shows an exponential relaxation approaching $V_{\Re} (\bx)$ at long time with a local relaxation time $\tau_{\bx} \equiv \frac{1}{2T}V_{\Im}(\bx)/V_{\Re}(\bx)$. 
This indicates that we can find signatures of the imaginary potential by measuring the relaxation behavior of the medium density long after the interaction quench. Notably, using Eq.~(\ref{eq:impot_asymp_long}) and the real-part potential found in Ref.~\cite{Fujii2022}, we find scaling behaviors of the local relaxation time as
\begin{align}
 \tau_{\bm{r}}
 \simeq  
 \frac{2\pi^3 T^3}{45 c_s^4} r^4  
\end{align}
at large distance for the superfluids. Similar scaling for $\tau_{\bm{r}}$ is also obtained for the Fermi case using Eq~(\ref{eq:ImV-asymptotic-Fermi-long}) although the large-distance Friedel oscillation at finite temperatures in $V_{\Re}(\br)$ does not admit as simple form as the superfluid one.

\section{Conclusion and outlook} \label{sec:conclusion}

We have studied the complex-valued in-medium potential between two heavy impurities immersed in quantum environments, with a particular focus on the Bose and Fermi polarons in cold atoms.
We provided a general definition of the in-medium potential \eqref{eq:def-complex-Vbar}-\eqref{eq:def-complex-V} from the long-time behavior of the correlation function for two immobile impurities immersed in the medium.
When the impurity-medium coupling is a weak $s$-wave contact coupling, the real and imaginary parts of the complex-valued potential are expressed by the low-frequency limit of the two-point retarded Green's functions of medium number density fluctuations as given in Eqs.~\eqref{eq:perturbative-V-Im}-\eqref{eq:retarded}.

We calculated the imaginary part in two setups relevant for cold atoms: the polarons in the free Fermi gas, and in the superfluid gas.
For the Fermi polaron case, using the fermionic field theory, we found the analytic formulae in Eqs.~\eqref{eq:V-imaginary-Fermi}-\eqref{eq:f-Fermi-polaron}. 
As shown in Fig.~\ref{fig:Imaginary-potential-Fermi}, it shows an oscillatory decay at low-temperature while the oscillation is suppressed at higher temperature.
In the case of polarons in the superfluid gas, using the superfluid effective field theory, we found the analytic formulae in Eqs.~\eqref{eq:impot1}-\eqref{eq:h-polaron-superfluid}, which shows a monotonic decay as shown in Fig.~\ref{fig:Imaginary-potential-Bose}.
Notably, the imaginary potential in both cases shows a power-law decay $V_{\mathrm{Im}}(r)\propto r^{-2}$ at long distance (see Figs.~\ref{fig:Imaginary-potential-Fermi} and \ref{fig:Imaginary-potential-Bose}), which originates not from the gapless nature of the medium excitations, but rather from the elastic scattering nature of the medium excitation with the heavy impurity.
We also proposed three experimental manifestations of the imaginary potential potentially accessible in cold-atomic systems---the RF interferometry signal, the  spectral width of bipolaron, and the medium density dynamics after a quench of the impurity.

Understanding the correlations between the polarons can lead to various interesting outlooks: with the imaginary part of the polaron-polaron potential, supplemented with the real part, we can formulate the full in-medium dynamics of the two polarons.
This will provide us with better understanding of the bipolaron problem, e.g., its binding, decoherence, lifetime, and dissipative quantum dynamics. 
The correlation between the polarons is also useful in understanding quantum phases in population-imbalanced quantum mixtures, such as the superfluid phase transitions~\cite{zwierlein2006fermionic,partridge2006pairing,PhysRevLett.97.190407}, or itinerant ferromagnetism~\cite{jo2009itinerant,PhysRevLett.118.083602}.

More generally, our work paves the way toward an open quantum system perspective of the polaron physics.
As the impurities are immersed in the quantum environment, the polaron physics should generally be a non-Hermitian open quantum problem.
The imaginary part essentially captures its non-Hermitian nature, such as the dissipative processes and decoherences of the impurities.
The complex-valued potential may seem to violate the impurity number conservation, but this can be resolved with a more sophisticated treatment of the stochastic potential~\cite{Akamatsu:2011se, Kajimoto:2017rel}, in which the imaginary potential is obtained from noise correlation.
By generalizing further to include the mobility of impurities, we can obtain additional velocity-dependent forces such as the drag force, which take into account the recoil of impurities.
With these generalizations, we may understand the polarons as quantum Brownian particles interacting with each other~\cite{Akamatsu:2020ypb}.

Our work is not only relevant for cold atoms, but also provides a universal understanding of the quantum impurity physics. 
The polaron phenomena not only appear in condensed matter physics and cold atoms, but a similar polaron-like phenomenon in subatomic physics has been experimentally realized in high-energy heavy-ion collisions, but its physical property has not yet been understood partly due to the complex and uncontrolled experimental situation.
For example, the dynamics of a bottomonium in the quark-gluon plasma (QGP) are analogous to the bipolaron problem~\cite{Akamatsu:2020ypb} and the dynamics of charm quarks and charmonia in the QGP are similar to the many-body problems of impurities~\cite{Du:2022uvj}. 
The universal aspect of the polaron physics explored in this work provides us with an interdisciplinary understanding of the polaron physics, ranging from QGP, condensed matter physics, and cold atoms. 
Such a universal understanding, together with a high degree of controllability in cold atoms, will lay the basis for building a quantum simulator of the polaron-like physics in QGP and condensed matter physics with cold atoms.

\section*{Acknowledgements} \label{sec:acknowledgements}
We thank Pak Hang Chris Lau and Tilman Enss for fruitful discussions.
This work is supported by Japan Society for the Promotion of Science (JSPS) KAKENHI Grant Number JP23H01174, and partially by RIKEN iTHEMS Program and YITP domestic molecule-style workshop at Kyoto University.
Y.A. is supported by JSPS KAKENHI Grant Number JP18K13538. 
S.E. is supported by JSPS KAKENHI Grant Numbers JP21H00116 and JP22K03492. 
K.F. is supported by the Deutsche Forschungsgemeinschaft (DFG, German Research Foundation), Project-ID 273811115 (SFB1225ISOQUANT) and JSPS KAKENHI Grant Number 24KJ0062.
M.H. is supported by JSPS KAKENHI Grant Number 22K20369.

\appendix
\begin{widetext}

\section{Derivation of the formula \eqref{eq:perturbative-V-Re}-\eqref{eq:imaginary-energy-disconnected} in the weak impurity-medium coupling case}
\label{sec:weak-coupling}

We here provide a detailed derivation of the perturbative formula \eqref{eq:perturbative-V-Re}-\eqref{eq:imaginary-energy-disconnected} for the complex-valued potential for the interested readers.
The derivation in this appendix serves as an altenative proof of those formulas based on the operator formalism rather than the path-integral formalism in Ref.~\cite{sighinolfi2022stochastic}.

For a system with weak impurity-medium coupling, defined by Eqs.~\eqref{eq:H-tot} and \eqref{eq:H-int} in the main text, the interaction picture with the choice of the unperturbed Hamiltonian 
\begin{align}
 \hH_0 = \hat{H}_{\mathrm{med}} +\hat{H}_{\mathrm{imp}} +g\int \diff^3 x \,\hat{\Phi}^\dagger(\bx)\hat{\Phi}(\bx)\average{\hat{n}(\bx)}, 
  \label{eq:def-H0}
\end{align}
gives a useful basis.
We then express the time-evolution operator $\hat U(t,0)=\rme^{-\rmi t\hat{H}_{\mathrm{total}}}$ as
\begin{align}
&\hat{U}(t,0) = \rme^{-\rmi t H_0} 
 \T \rme^{
    -\rmi g \int^{t}_{0} \diff t^{\prime} \diff^3x\,
    \hat{\Phi}^\dagger(\bx,t')\hat{\Phi}(\bx,t')
    \delta\hat{n}(\bx,t')},
 \label{eq:U-reorganized}
\end{align}
with $\T$ denoting the time-ordered product.
Here, we introduced $\hat{\Ocal} (t) \equiv \rme^{\rmi t\hat{H}_0} \hat{\Ocal} \rme^{-\rmi t\hat{H}_0}$ for an arbitrary operator $\hat{\Ocal}$, and the number density fluctuation operator around its thermal average as $\delta\hat{n}(\bx)=\hat{n}(\bx)-\average{\hat{n}(\bx)}$.

We take the heavy-impurity limit as the first step to evaluate $\Psi(\br,t)$.
In this limit, the large impurity mass $M_*^{-1} \simeq M^{-1}$ suppresses its kinetic terms, i.e. the impurity is immobile and can be treated as a test particle fixed at a certain position.
Then, the impurity density operator $\hPhi^\dag (\bx,t) \hPhi (\bx,t)$ is time-independent and we obtain
\begin{align}
 &\Psi(\br,t)
 = \rme^{-\rmi tg(\average{\hat{n}(\bx_1)}+\average{\hat{n}(\bx_2)})} \left\langle
    \T \exp\biggl[
        - \rmi g\int_0^t \diff t'
        \,\Bigl(\delta\hat{n}(\bx_1,t')+\delta\hat{n}(\bx_2,t')\Bigr)
    \biggr]
 \right\rangle\Psi(\br,0).
\end{align}

To proceed further, we employ the weak-coupling expansion with respect to the impurity-medium coupling $g$.
In the ladder approximation, we get
\begin{align}
 &\Psi(\br,t)
 \simeq \rme^{-2\rmi tg\average{\hat{n}(\bzero)}}
 \exp\biggl[
  -g^2\int^{t}_0 \diff t_1 \int^{t}_0 \diff t_2\,
  \Big(
  \langle \T \delta\hat{n}(\bm 0,t_1)\delta\hat{n}(\bm 0,t_2)\rangle 
  + \langle \T \delta\hat{n}(\bm r,t_1)\delta\hat{n}(\bm 0,t_2)\rangle
  \Big)
\biggr]\Psi(\br,0),
\end{align}
where the translational and rotational invariance of the medium is used.
Matching this expression with the aforementioned long-time behavior $\Psi(\br,t) \simeq \rme^{-\rmi \bar{V}(\br)t}\Psi(\br,0)$, we find $\bar{V}(\br)$ at $O(g^2)$ as 
\begin{align}
&\bar{V}(\br) 
= 2g\average{\hat{n}(\bzero)} 
+ \lim_{t \to \infty}\frac{g^2}{\rmi t}
\int_0^{t} \diff t_1 \int^{t}_{0} \diff t_2\, 
\Big(
\langle \T \delta\hat{n}(\bm 0,t_1)\delta\hat{n}(\bm 0,t_2)\rangle 
 + \langle \T \delta\hat{n}(\bm r,t_1)\delta\hat{n}(\bm 0,t_2)\rangle
\Big)
+ O(g^3).
\label{eq:Vbar-computation}
\end{align}
According to Eq.~\eqref{eq:def-complex-V}, subtracting the one-body energy, we arrive at the expression of the in-medium interaction potential $V(\br)$ for a weak impurity-medium coupling case as
\begin{align}
 V (\br)
 &\simeq 
 \lim_{t \to \infty} \frac{g^2}{\rmi t}
 \int_0^{t} \diff t_1 \hspace{-2pt} 
 \int_0^t \diff t_2\,
 \langle \T \delta\hat{n}(\bm r,t_1)\delta\hat{n}(\bm 0,t_2)\rangle,
 \nonumber \\
 \cout{
 &\rsout{= \lim_{t \to \infty} \frac{2g^2}{\rmi t}
 \int_0^{t} \diff s\, (t-s)
  \langle \delta\hat{n}(\bm r,s)\delta\hat{n}(\bm 0,0)\rangle
 \nonumber } \\ 
 }
 &=- 2 \rmi g^2 
 \int_0^{\infty} \diff s\,  \langle \delta\hat{n}(\bm r,s)\delta\hat{n}(\bm 0,0)\rangle,
 \label{eq:perturbative-V}
\end{align}
where we dropped $O(g^3)$ terms and assumed that $\langle \delta\hat{n}(\bm r,s)\delta\hat{n}(\bm 0,0)\rangle$ at fixed $\br$ decays in a finite duration of time.
We here note that the zero temperature limit of Eq.~\eqref{eq:perturbative-V}, expressed in terms of the time-ordered product, reproduces a familiar formula of the induced potential at weak coupling~\cite{fetter2012quantum}.

By separating the real and imaginary parts of Eq.~\eqref{eq:perturbative-V}, we can express them with the retarded Green's function.
For instance, recalling $\average{\hat{A} \hat{B}}^* = \average{\hat{B}^\dag \hat{A}^\dag }$ and using the fluctuation-dissipation relation, we find
\begin{align}
V_{\mathrm{Im}} (\br) 
  &= - \frac{g^2}{2} \hspace{-3pt} 
 \int_{-\infty}^{\infty} \hspace{-7pt} \diff s\,  
 \Big\langle
  \delta\hat{n}(\bm r,s)\delta\hat{n}(\bm 0,0)
 + \delta\hat{n}(\bm 0,0) \delta\hat{n}(\bm r,s)
 \Big\rangle
 = - g^2 \lim_{\omega \to 0} \frac{2T}{\omega} \Im G^R (\br,\omega) .
\end{align}
This completes the derivation of the perturbative formulas~\eqref{eq:perturbative-V-Re}-\eqref{eq:imaginary-energy-disconnected} in the main text.

\section{Complex potential in open systems} \label{sec:appendix2}

To our knowledge, the in-medium complex potential for heavy impurities was first applied to the cold atomic physics in Ref.~\cite{sighinolfi2022stochastic}, but the main subject was classical Langevin equation.
In this appendix, we explain how our work is related to the formulation given in Ref.~\cite{sighinolfi2022stochastic}, by presenting relations between the complex potential and several concepts in quantum and classical open systems such as influence functional, decoherence, and Langevin equation.
From the perspective of open system, we gain a comprehensive understanding of the complex potential, which allows us to calculate physical quantities beyond the spectral density of impurities.

The reduced density matrix of $N$ impurities, $\rho(\bm Q, \bm Q',t)$ with $\bm Q = (\bm q_1, \bm q_2, \cdots, \bm q_N)$, evolves from an initial condition $\rho(\bm Q_I, \bm Q_I',t_I)$ to a final one $\rho(\bm Q_F, \bm Q_F',t_F)$ by an evolution kernel $\mathcal K(\bm Q_F, \bm Q_F',t_F|\bm Q_I, \bm Q_I',t_I)$
\begin{align}
 \rho(\bm Q_F, \bm Q_F',t_F) 
 = \int \diff \bm Q_I \diff \bm Q_I'\mathcal K(\bm Q_F, \bm Q_F',t_F|\bm Q_I, \bm Q_I',t_I)
 \rho(\bm Q_I, \bm Q_I',t_I),
\end{align}
where the kernel $\mathcal K$ is given by
\begin{align}
\mathcal K(\bm Q_F, \bm Q_F',t_F|\bm Q_I, \bm Q_I',t_I) 
 =\int_{\bm Q_I, \bm Q_I'}^{\bm Q_F, \bm Q_F'} {\mathcal D}\bm Q{\mathcal D}\bm Q'
 \exp 
 \left( 
 \rmi \int \diff t 
 \frac{M}{2}\sum_j (\dot{\bm q}_j^2 - \dot{\bm q}_j^{\prime 2}) 
 + \rmi \Phi[\bm Q,\bm Q']
 \right).
\end{align}
Here $\Phi[\bm Q,\bm Q']$ is called an influence functional~\cite{Feynman:1963fq,Caldeira1983}.
When the impurity and medium are coupled weakly by the contact interaction \eqref{eq:H-int}, the influence functional is given in Ref.~\cite{sighinolfi2022stochastic}
\begin{align}
 \Phi[\bm Q,\bm Q']
 &= -\frac{1}{2}\int_{t_I}^{t_F} \diff t \sum_{i,j} V_{\Re}(\bm q_i-\bm q_j) - V_{\Re}(\bm q_i'-\bm q_j') 
 - \frac{\rmi}{2}
 \int_{t_I}^{t_F} \diff t\sum_{i,j}
 \left(
 \begin{aligned}
 &V_{\Im}(\bm q_i-\bm q_j) - V_{\Im}(\bm q_i-\bm q_j') \\
 &- V_{\Im}(\bm q_i'-\bm q_j) + V_{\Im}(\bm q_i'-\bm q_j')
\end{aligned}\right)
 + O(\dot Q, \dot Q'), 
\end{align}
where we use our notation for the medium correlation functions.
The higher order terms in the derivative expansion are omitted for simplicity.
From this Markovian influence functional, one can derive quantum master equation for $N$ impurities.
In Ref.~\cite{sighinolfi2022stochastic}, the authors take a classical limit to obtain Fokker-Planck equation, from which they further get generalized Langevin equation for the impurities.
For details, we refer to \cite{Blaizot:2015hya} as well as \cite{sighinolfi2022stochastic}, latter of which uses the formalism developed in the former.

In the influence functional in the recoilless limit, i.e. ignoring $O(\dot Q, \dot Q')$ terms, the imaginary potential can be expressed as an average over random phases
\begin{align}
 &\exp\left[
 \frac{1}{2}
 \int_{t_I}^{t_F} \diff t\sum_{i,j}
 \left(
 \begin{aligned}
   &V_{\Im}(\bm q_i-\bm q_j) - V_{\Im}(\bm q_i-\bm q_j') \\
    &- V_{\Im}(\bm q_i'-\bm q_j) + V_{\Im}(\bm q_i'-\bm q_j')
 \end{aligned}
 \right)\right] 
= \left\langle\exp
\left[- \rmi\int_{t_I}^{t_F} \diff t 
  \sum_{i}\theta(\bm q_i,t) - \theta(\bm q_i',t)\right]\right\rangle_{\theta},
\end{align}
where $\theta(\bm x,t)$ is a Gaussian white noise with finite correlation length [note that $-V_{\Im}(\bm x)$ is positive definite] satisfying
\begin{align}
    \langle\theta(\bm x,t)\theta(\bm x',t')\rangle_{\theta} = -V_{\Im}(\bm x-\bm x')\delta(t-t').
\end{align}
Therefore, in the recoilless limit, we can formulate the impurity dynamics by a Hamiltonian with stochastic potential
\begin{align}
    H_{\theta}=\sum_i\frac{\bm p_i^2}{2M} + \frac{1}{2}\sum_{i,j}V_{\Re}(\bm q_i-\bm q_j) + \sum_i\theta(\bm q_i,t).
\end{align}
The correlation length of $\theta(\bm x,t)$ determines the effectiveness of wave function decoherence.
This stochastic picture enables us to intuitively understand the origin of the imaginary part, which is derived using $\Psi(\bm r,t)$ in Eq.~\eqref{eq:impurity-Green-fcn}.
The real-time correlation function $\Psi(\bm r,t)$ is actually an averaged wave function evolved in the presence of the noise and partial cancellation of the amplitudes by the random phases results in the imaginary potential.
It also provides a simple derivation of the generalized Langevin equation~\cite{sighinolfi2022stochastic} with the help of fluctuation-dissipation relation.
Since the random force at the position of a classical particle $\bm r_i$ is given by the gradient of the stochastic potential there
\begin{align}
    \bm f(\bm r_i,t) = -\bm\nabla \theta(\bm r_i, t),
\end{align}
the correlation of random force is
\begin{align}
    \langle f_{\alpha}(\bm r_i,t)f_{\beta}(\bm r_j,t')\rangle_f
    = \partial_{\alpha}\partial_{\beta} V_{\Im}(\bm r_i-\bm r_j)\delta(t-t').
\end{align}
The dissipative force is determined by the fluctuation-dissipation relation.
Note that the universal power-law decay of $V_{\Im}(r)\propto 1/r^2$ at long distance suggests the random force correlation decays $\propto 1/r^4$ at distant points.

Finally, let us apply the stochastic potential to the calculation of transition rate $\gamma$ of a bound state $\Psi_b$.
To calculate it, let us first calculate the survival probability $c(t)$ after a short time period $t=\Delta t$
\begin{align}
 c(\Delta t) = 
 \left\langle \left|
  \int \diff^3 r \Psi_b^*(\bm r) \rme^{-\rmi\Delta t H_{\theta}}\Psi_b(\bm r)
 \right|^2\right\rangle_{\theta}.
\end{align}
After some algebra, we can get the transition rate 
\begin{align}
 \gamma &\equiv -\dot c(t=0) 
 =\Gamma
 +2\int \diff^3 r \diff^3r' |\Psi_b(\bm r)|^2 |\Psi_b(\bm r')|^2
 \left[V_{\Im}(\bm r_+) + V_{\Im}(\bm r_-)\right],
\end{align}
where $\bm r_{\pm}=\frac{1}{2}(\bm r\pm \bm r')$ and $\Gamma$ is given in Eq.~\eqref{eq:width}.
The second term subtracts from the total width $\Gamma$ the rate of staying in $\Psi_b$ after a collision, which approaches $-\Gamma$ when the bound state size is small.
Therefore, the transition rate is actually smaller than the spectral width.
Similar calculation was done in quarkonium in the quark-gluon plasma~\cite{Kajimoto:2017rel}, but the total width and the transition rate are equal (i.e., the second term vanishes) because of the opposite charges of the heavy quark-antiquark pair.

\section{Derivation of Eqs.~\eqref{eq:impot1}-\eqref{eq:impot_asymp_long}
} \label{sec:appendix1}

In this appendix, we provide a derivation of equations \eqref{eq:impot1}-\eqref{eq:impot_asymp_long} presented in the main text.
We begin with the equation \eqref{eq:impot-superfluid}, and for clarity, we will present it again here as:
\begin{align}
 V_{\Im}(\bm r) &= 
 - \frac{2\pi g^2}{m^2} \int_{\bk,\bq} 
 \rme^{\rmi(\bk-\bq) \cdot \br} 
 \frac{(\bq\cdot\bk)^2}{4 E_{\bk}^2}
 \delta ( E_{\bk}- E_{\bq})
 [1+n_B(E_{\bk})] n_B (E_{\bk}) .
\end{align}
We first decompose the momentum integral into the angular parts as 
\begin{align}
 V_{\Im} (\br) 
 &= - \frac{2\pi g^2}{m^2} 
 \left( \frac{1}{2\pi^2} \right)^2
 \int_0^{\infty} \diff k k^2 
 \int_0^\infty \diff q q^2
 \int \frac{\diff \Omega_{\bk}}{4\pi} \frac{\diff \Omega_{\bq}}{4\pi} 
 \rme^{\rmi (\bk - \bq) \cdot \br}
 \frac{ (\bk\cdot\bq)^2}{4c_s^2 k^2}
 \frac{1}{c_s} \delta(k-q) 
 [1+n_B(c_sk)] n_B(c_sk) \nonumber \\
 &= - \frac{g^2}{8\pi^3 m^2 c_s^3} 
 \int_0^{\infty} \diff k  \int_0^\infty \diff q q^2
 \delta(k-q) [1+n_B(c_sk)] n_B(c_sk)
 \int \frac{\diff \Omega_{\bk}}{4\pi} \frac{\diff \Omega_{\bq}}{4\pi}
 \rme^{\rmi (\bk - \bq) \cdot \br} (\bk\cdot\bq)^2 ,
\end{align}
where $\diff \Omega_{\bk}$ denotes the angular integral for the momentum $\bk$.
We can perform the angular integrals as
\begin{align}
 \int \frac{\diff \Omega_{\bk}}{4\pi} \frac{\diff \Omega_{\bq}}{4\pi}
 \rme^{\rmi (\bk - \bq) \cdot \br} (\bk\cdot\bq)^2  
 &= 
 \left(  
  \frac{\partial^2}{\partial r_i \partial r_j}
  \int \frac{\diff \Omega_{\bk}}{4\pi} \rme^{\rmi \bk \cdot \br}
 \right)
 \left(  
  \frac{\partial^2}{\partial r_i \partial r_j}
   \int \frac{\diff \Omega_{\bq}}{4\pi} \rme^{-\rmi \bq \cdot \br}
 \right)
 \nonumber \\
 &= 
 \left(  
  \frac{\partial^2}{\partial r_i \partial r_j} \frac{\sin kr}{kr}
 \right)
 \left(  
  \frac{\partial^2}{\partial r_i \partial r_j}  \frac{\sin qr}{qr}
 \right),
\end{align}
where summation over repeated indices $(i,j)$ is implied.
After performing the $q$-integral with the help of the delta function, we obtain
\begin{align}
 V_{\Im} (\br) 
 &= - \frac{g^2}{8\pi^3 m^2 c_s^3} 
 \int_0^{\infty} \diff k k^6 [1+n_B(c_sk)] n_B(c_sk)
 \left.
 \left(  
  \frac{\partial^2}{\partial z_i \partial z_j} \frac{\sin z}{z}
 \right)
 \left(  
  \frac{\partial^2}{\partial z_i \partial z_j}  \frac{\sin z}{z}
 \right)
 \right|_{\bz=k\br} ,
 \label{eq:ImV-sinz-z}
\end{align}
To proceed further, it is crucial to recall the formula for the spherical Bessel functions
\begin{align}
 j_0 (z) = \frac{\sin z}{z}, \quad
 \left(\frac{\diff}{\diff z} - \frac{n}{z}\right) j_n(z) = -j_{n+1}(z),
\end{align}
which enables us to obtain
\begin{align}
 \frac{\partial^2}{\partial z_i \partial z_j}\frac{\sin z}{z}
 =\frac{\partial^2}{\partial z_i \partial z_j} j_0(z)
&=-\delta_{ij}\frac{j_1(z)}{z} +\frac{z_i z_j}{z^2} j_2(z), 
\\
\left(\frac{\partial^2}{\partial z_i\partial z_j}\frac{\sin z}{z}\right)
\left(\frac{\partial^2}{\partial z_i\partial z_j}\frac{\sin z}{z}\right)
&= \frac{3 j_1(z)^2}{z^2} - \frac{2j_1(z) j_2(z)}{z} + j_2(z)^2.
\end{align}
Substituting this result into Eq.~\eqref{eq:ImV-sinz-z}, we obtain
\begin{align}
 V_{\Im} (\br) 
 &= - \frac{g^2}{8\pi^3 m^2 c_s^3} 
 \int_0^{\infty} \diff k k^6 [1+n_B(c_sk)] n_B(c_sk)
 \left[
  \frac{3j_1(kr)^2}{(kr)^2}- \frac{2j_1(kr)j_2(kr)}{kr} + j_2(kr)^2
 \right]
\end{align}
To make the integration variable dimensionless, we define $s \equiv c_sk/T = k/k_T$, which results in Eq.~\eqref{eq:impot1}:
\begin{align}
 V_{\Im}(r)
 &=-\frac{g^2 T^7}{8\pi^3 m^2c_s^{10}} \int_0^{\infty} \diff s 
 \frac{s^6 \rme^{s}}{(\rme^{s}-1)^2}
 \left[\frac{3j_1(s k_Tr)^2}{(s k_Tr)^2}- \frac{2j_1(s k_Tr)j_2(s k_Tr)}{s k_Tr} + j_2(s k_Tr)^2\right] \nonumber\\
 &=-\frac{g^2 T^7}{8\pi^3 m^2c_s^{10}}  h(k_T r)
 \label{eq:ImV-appendix}
\end{align}
with the function $h(y)$ defined as in Eq.\eqref{eq:h-polaron-superfluid}.
Moreover, the asymptotic behaviors of the spherical Bessel functions
\begin{align}
 j_n(x) &\sim 
 \begin{cases}
 \dfrac{x^n}{(2n+1)!!} \quad (x\to 0),
 \vspace{5pt} \\
 \dfrac{1}{x}\sin\left(x-\dfrac{n\pi}{2}\right) \quad (x\to \infty),
 \end{cases}
\end{align}
allows us to derive $h(y)$ at $y=0$ and $y\gg 1$ as
\begin{align}
h(y = 0)
&= \int_0^{\infty} \diff s\frac{s^6 \rme^{s}}{(\rme^{s}-1)^2}\frac{1}{3}
=\frac{16\pi^6}{63} , \\
h(y \gg 1)
&\simeq \int_0^{\infty} \diff s \frac{s^6 \rme^{s}}{(e^{s}-1)^2}
\frac{\sin^2(sy)}{(sy)^2}
\simeq  \frac{1}{y^2} \int_0^{\infty} ds\frac{s^4\rme^{s}}{(\rme^{s}-1)^2}\frac{1}{2}
= \frac{2\pi^4}{15y^2}.
\end{align}
Substituting these results into Eq.~\eqref{eq:ImV-appendix}, we obtain the asymptotic form of the imaginary potential given in Eqs.~\eqref{eq:impot_asymp_short}-\eqref{eq:impot_asymp_long} in the main text.

\end{widetext}

\bibliography{Refs}

\end{document}